\documentstyle[dina4,epsf,twocolumn]{article}
\begin{document}
\renewcommand{\textfraction}{0.0}
\renewcommand{\topfraction}{1.0}
\renewcommand{\bottomfraction}{1.0}
\newcommand{\gapro}
  {\raisebox{-0.25ex} {$\,\stackrel{\scriptscriptstyle>}%
    {\scriptscriptstyle\sim}\,$}}
\newcommand{\lapro}
   {\raisebox{-0.25ex} {$\,\stackrel{\scriptscriptstyle<}%
    {\scriptscriptstyle\sim}\,$}}
\def\cH{{\cal{H}}}
\def\om{\omega_0}
\def\bid{b_i^\dagger}
\def\bi{b_i^{}}
\def\ep{\varepsilon_p}
\def\figurename{Fig.}
\rule[-8mm]{0mm}{8mm}
\begin{minipage}[t]{16cm}
{\large \bf  SPECTRAL PROPERTIES OF THE 2D HOLSTEIN t--J MODEL\\[4mm]}
H.~Fehske$^{1}$, G.~Wellein$^{1}$, B.~B\"auml$^{1}$
and R.~N.~Silver$^{2}$,\\[3mm]
$^1$Physikalisches Institut, Universit\"at Bayreuth, D--95440 Bayreuth,
Germany\\
$^{2}$MS B262 Los Alamos National Laboratory, Los Alamos,
NM 87545, USA
\\[4.5mm]
\hspace*{0.5cm}
Employing the Lanczos algorithm in combination with 
a kernel polynomial moment expansion (KPM) 
and the maximum entropy method (MEM), 
we show a way of calculating charge and spin excitations 
in the Holstein t--J model, including
the full quantum nature of phonons. 
To analyze polaron band formation we evaluate 
the hole spectral function for a wide range of 
electron--phonon coupling strengths. 
For the first time, we  present results for the optical conductivity 
of the 2D Holstein t--J model. 
\end{minipage}\\[4.5mm]
\normalsize
Polaronic features of dopant--induced charge carriers have been
observed in the isostructural copper--based and nickel--based 
charge--transfer oxides $\rm La_{2-x}Sr_x[Cu,Ni]O_{4+y}$~[1]. 

Studying (bi)polaron effects in such strongly coupled electron--phonon (EP)
systems, the Holstein t--J model (HtJM) 
has recently attracted much attention~[2]. 
The HtJM Hamiltonian reads 
\begin{eqnarray}
\cH\!\!\!\!&=&\!\!\!\! \hbar\om \sum_i \Big(\bid\bi + \mbox{\small $\frac{1}{2}$}\Big) 
\!-\! \sqrt{\ep\hbar\om}  \sum_i \big(\bid + \bi \big)\,\tilde{h}_i^{}\nonumber
\\&&\!\! -t\sum_{\langle i,j \rangle \sigma} 
\tilde{c}_{i\sigma}^\dagger 
\tilde{c}_{j\sigma}^{} +J \sum_{\langle i j\rangle}
\Big(\vec{S}_i^{}\vec{S}_j^{} -
\frac{\tilde{n}_i^{}\tilde{n}_j^{}}{4} \Big)\,.
\end{eqnarray}
The first two terms take into account the phonon part 
and the EP interaction, respectively, whereas the last two terms 
represent the standard \mbox{t--J} model acting in a Hilbert space 
without double occupancy. In~(1), doped holes ($\tilde{h}_i=1-\tilde{n}_i$) 
are coupled locally to a dispersionsless optical phonon mode ($\ep$ -
EP coupling constant, $\om$ -- bare phonon frequency).

In this contribution, we investigate the HtJM by performing 
exact diagonalizations on a square ten--site 
lattice, where the phonon degrees of freedom are treated 
within a well--controlled Hilbert space truncation procedure~[3]. 
To obtain information about dynamical properties 
of the model under consideration, we combine the 
Lanczos algorithm with the KPM and MEM approaches~[4]. 

In order to address the problem of polaron formation
in an antiferromagnetic correlated spin background,    
\begin{figure}
\rule[-5.cm]{0mm}{5.4cm}
\end{figure}
\begin{figure}[b]
\vspace*{-0.6cm}
\centerline{\mbox{\epsfxsize 7.4cm\epsffile{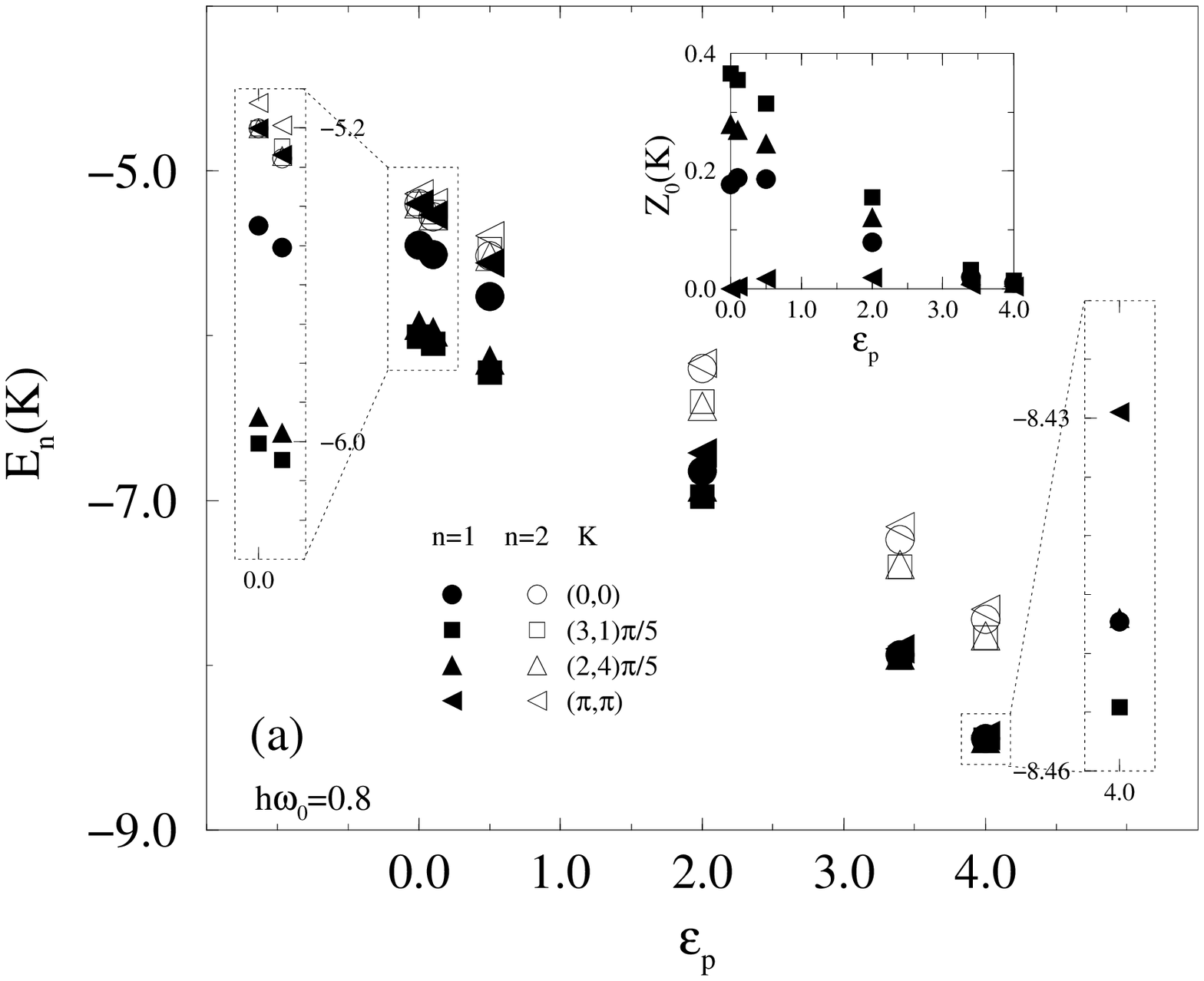}}}
\vspace*{-0.5cm}
\centerline{\mbox{\epsfxsize 7.4cm\epsffile{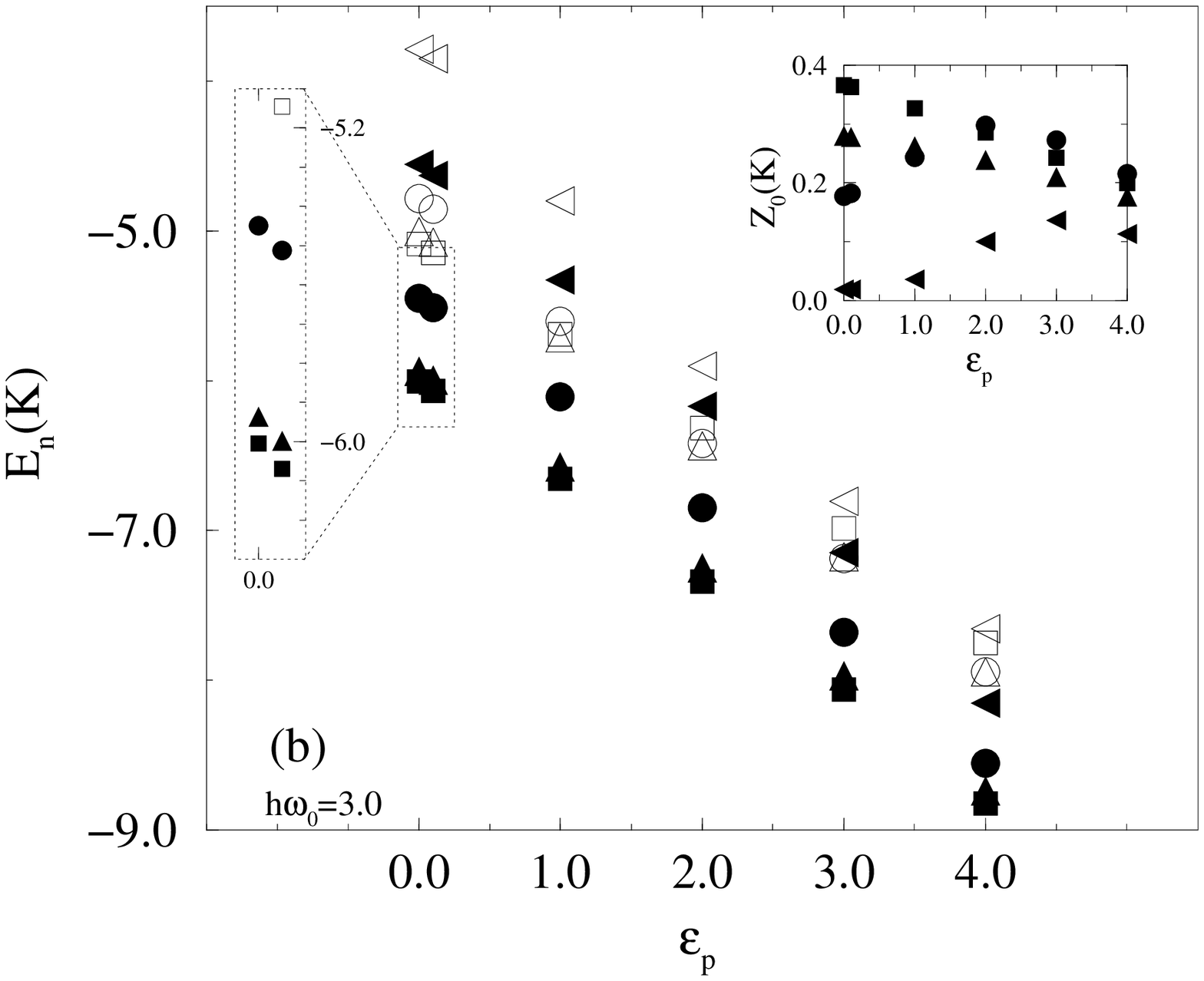}}}
\vspace*{-0.4cm}
\caption{Polaron band formation in the 2D HtJM. The    
wave function renormalization factors, $Z_0(\vec{K})\propto
\sum_\sigma |\langle {\mit \Psi}_0^{(N-1)}(\vec{K})
|\tilde{c}_{-\vec{K},\sigma}|
 {\mit \Psi}_0^{(N)}(\vec{0})\rangle |^2$,  
are given as a function of $\ep$ in the insets.}
\end{figure}
we have calculated the $\vec{K}$--resolved spectral function $A_{\vec{K}}(E)$ 
for a single dynamical hole at $J=0.4$ (energies in units of $t$)~[5]. 
The positions of the two lowest peaks of $A_{\vec{K}}(E)$,
denoted by $E_{0/1}(\vec{K})$, 
are displayed as a function of EP coupling strength in Fig.~1 
for the allowed $\vec{K}$ vectors at $\hbar\om=0.8$ and 3.0. 
As expected, in the very weak  EP coupling limit 
the low--lying excitations are 
t--J hole--quasiparticles  weakly dressed by phonons. 
For low and intermediate phonon frequencies, 
the energy to excite one phonon lies inside 
the quasiparticle band of the pure t--J model. Thus 
at arbitrarily small $\ep$, predominantly phononic states 
with a small admixture of electronic character enter the low--energy
spectrum in {\it all} $\vec{K}$--sectors [cf. the region about 
$E_n\simeq -5.2$ in the inset of (a)].   
\begin{figure}[t]
\begin{minipage}[b]{16cm}
\centerline{\mbox{\epsfxsize 5.3cm\epsffile{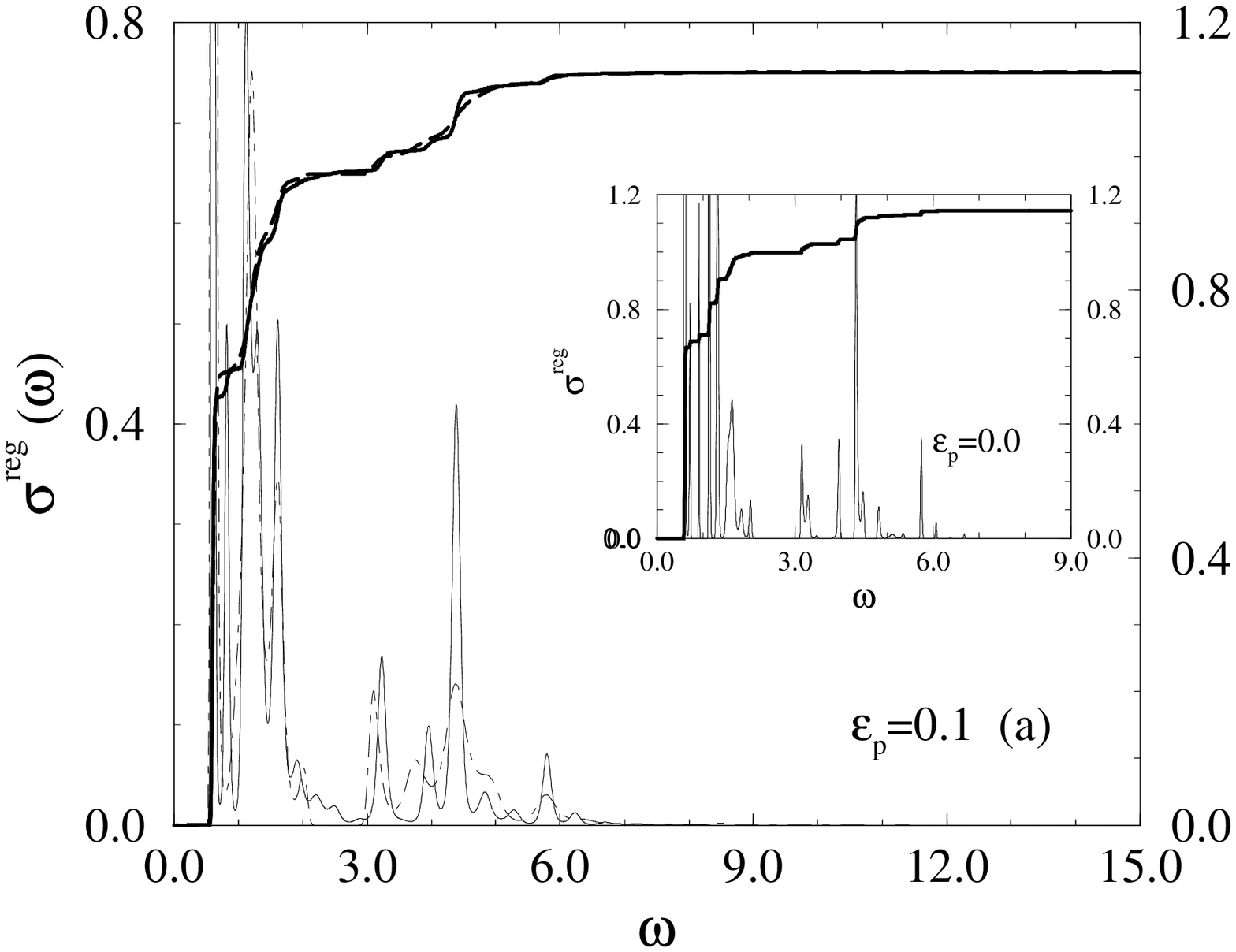}}
 \hspace*{-0.3cm}      \mbox{\epsfxsize 5.3cm\epsffile{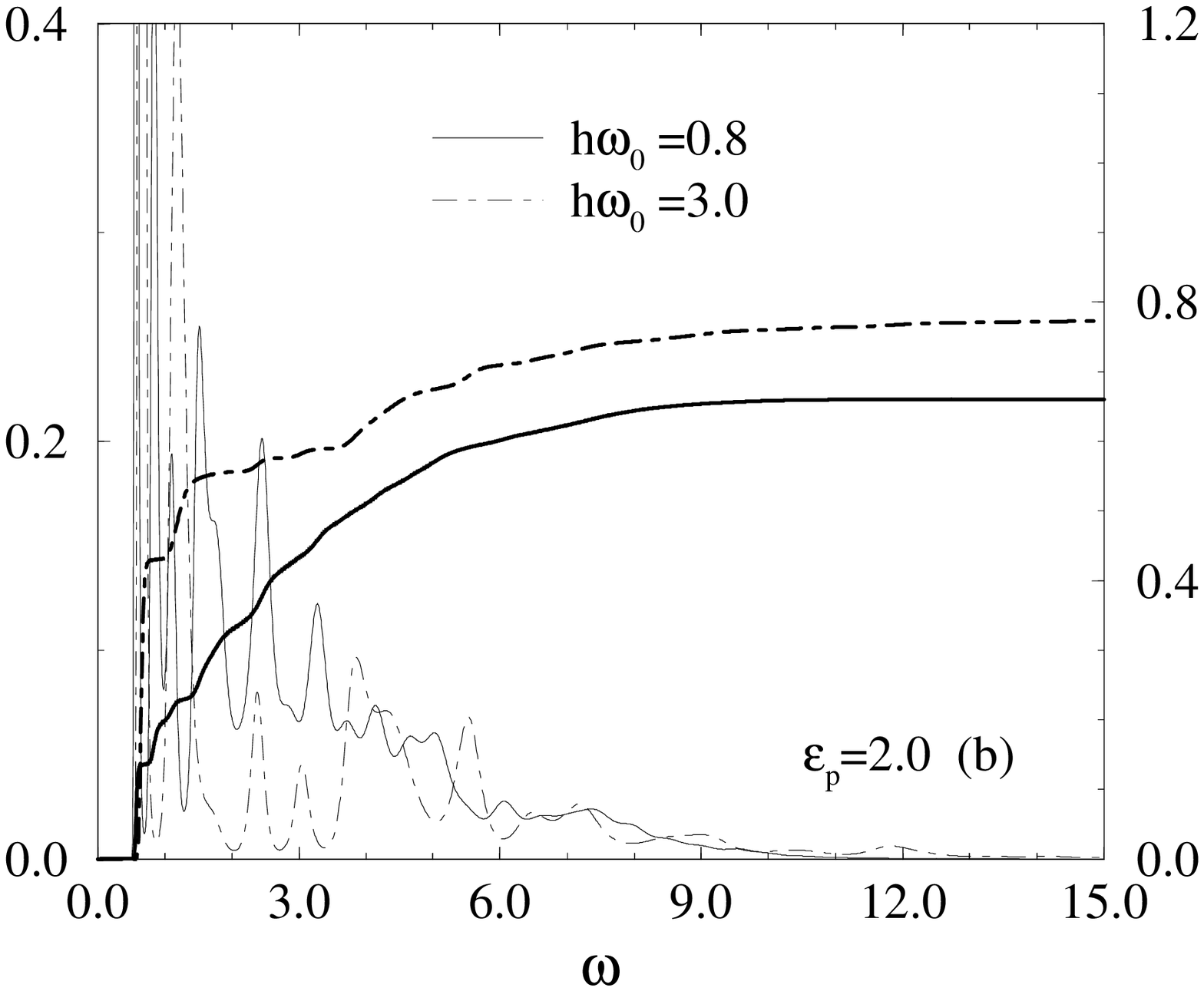}}
  \hspace*{-0.3cm}      \mbox{\epsfxsize 5.3cm\epsffile{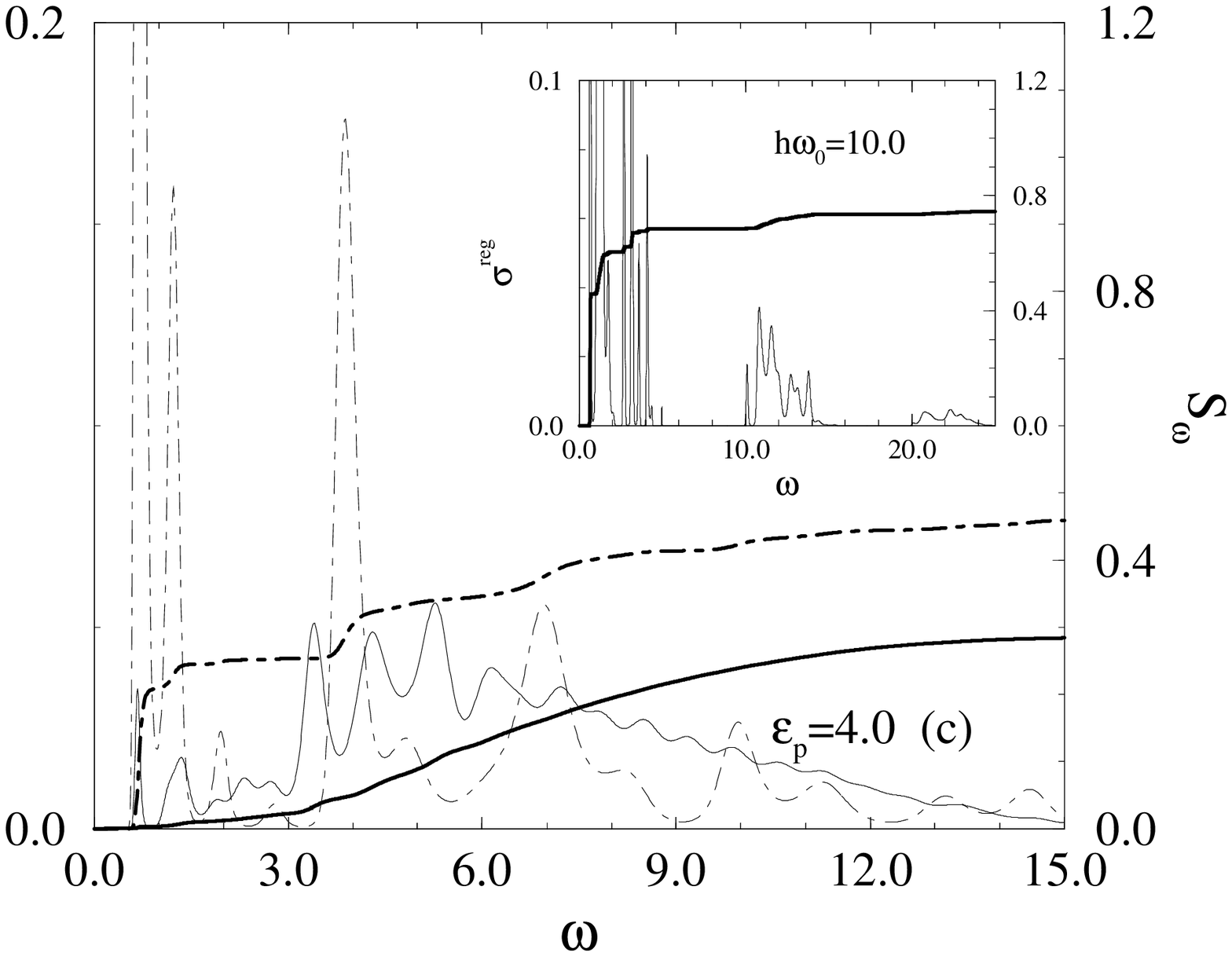}}}
\vspace*{-0.3cm}
\caption{Optical conductivity $\sigma_{xx}^{reg}(\omega)$ 
and $\omega$--integrated spectral weight in the dissipative part of 
$\sigma^{reg}$,  $S_{\omega}=\int_0^\omega d\omega^\prime 
\sigma_{xx}^{reg}(\omega^\prime)$, for the 2D single--hole HtJM 
with 12 phonons (periodic boundary conditions).}
\end{minipage}\\[-.9cm]
\end{figure}
With increasing $\ep$ a strong mixing of holes and phonons 
takes place, whereby both quantum objects completely 
loose their own identity, and finally an extremely narrow 
well--separated {\it polaron} band is formed at large $\ep$. 
This scenario is corroborated by the behavior of
the $\vec{K}$--dependent renormalization factor $Z_0(\vec{K})$ shown in the
upper insets, which 
can be taken as a measure of the ``electronic'' contribution to 
the polaronic quasiparticle (see insets). As can be seen from
 Fig.~1~(b), the phonon induced band 
renormalization is weakened in the non--adiabatic regime, 
where retardation and multi--phonon effects are of minor importantance.

To discuss the influence of the EP coupling 
on the optical response of the system,  
let us evaluate the regular part of the optical conductivity 
at finite energy transfer $\omega $,
\begin{equation}
\sigma_{xx}^{reg}(\omega)=\frac{e^2\pi}{N}\sum_{n\neq 0}
\frac{|\langle {\mit \Psi}_n|\hat{j}_x| {\mit \Psi}_0\rangle |^2}
{E_n-E_0}\;\delta [\omega-(E_n-E_0)]\,.
\end{equation}
Results for $\sigma_{xx}^{reg}(\omega)$ are presented 
in Fig.~2 for $\ep=0.1$ (a), 2 (b), and 4 (c).
In the weak EP coupling region, we recover the main features of the 
optical conductivity of the t--J model (inset Fig~2.~(a)), i.e., (i) 
an ``anomalous'' broad mid--infrared absorption band 
[$J\lapro\omega\lapro 2t$], 
separated from the Drude peak [$D\,\delta (\omega)$; not shown] 
by a ``pseudo-gap'' $\sim J$, and (ii) an  ``incoherent'' 
tail up to $\omega\sim 7t$. At  
larger $\ep$, we observe a redistribution of spectral weight 
to higher energies (cf.~${\mit \Delta} S_\omega$), which is 
much more pronounced in the {\it adiabatic} regime. 
In particular, the transition to the  (hole) polaron state 
is accompanied by the development of a broad maximum in  
$\sigma_{xx}^{reg}(\omega)$ at $\omega\lapro2\ep$, whereas
the optical response becomes strongly suppressed 
at low $\omega$. Most notably, $\sigma_{xx}^{reg}(\omega)$ 
has an highly {\it asymmetric} lineshape at 
intermediate frequencies and coupling strengths 
as observed, e.g., for  $\rm La_{0.9}Sr_{0.1}NiO_{4}$~[1]. 
\begin{figure}[t]
\rule[-5.cm]{0mm}{2.8cm}
\end{figure}
This effect can be traced back to a rather broad ground--state 
phonon distribution function obtained for $\ep=2, 4$ and $\hbar \om=0.8$~[5]. 
Contrary, in the anti--adiabatic limit, the ``electronic'' 
lineshape is much less affected, but $\sigma_{xx}^{reg}(\omega)$ shows 
additional superstructures corresponding to ``interband'' transitions 
between t--J--like absorption bands with different 
number of phonons [see Fig~2~(c), inset].
These results clearly demonstrate the 
complex interplay of electron and EP correlation effects. 
\\[0.1cm]
REFERENCES
\begin{enumerate}
\vspace*{-0.2cm}
\item X.-X. Bi and P. C. Eklund, Phys. Rev. Lett. {\bf 70},
2625 (1993). 
\vspace*{-0.2cm}
\item 
H.$\!$ Fehske$\!$ {\it et al.},
Phys.$\!$ Rev.$\!$ B$\!$ {\bf 51},$\!$ 16582$\!$ (1995);
A.$\!$ Dobry$\!$ {\it et al.}, Phys.$\!$ Rev.$\!$ B$\!$ {\bf 52},$\!$ 13722$\!$ (1995).\vspace*{-0.2cm} 
\item G. Wellein, H. R\"oder, and H. Fehske, Phys. Rev. B {\bf 53}, 
9666 (1996).
\vspace*{-0.2cm}
\item R. N. Silver, et al., J. of Comp. Phys. {\bf 124}, 115 (1996).
\vspace*{-0.2cm}
\item H. Fehske$\!$ {\it et al.}, Physica B, to appear (1997).
\end{enumerate}
\end{document}